\documentclass[11pt,letterpaper,final]{JHEP3}

\usepackage{amsfonts,amsmath,amssymb}
\usepackage{bm,bbm}
\usepackage{graphicx,epsfig}

\newcommand{\gsim}{\mathrel{\lower0.8ex\vbox{\lineskip=0.15ex\baselineskip=0ex
                   \hbox{$>$}\hbox{$\sim$}}}}
\newcommand{\lsim}{\mathrel{\lower0.8ex\vbox{\lineskip=0.15pt\baselineskip=0ex
                   \hbox{$<$}\hbox{$\sim$}}}}

\newcommand{\braket}[2]{\left\langle #1 | #2 \right\rangle}  
\newcommand{\sla}[1]{{\raise.15ex\hbox{$/$}\kern-.57em #1}}
\newcommand{\Sla}[1]{\kern0.12em{\raise.15ex\hbox{$/$}\kern-.74em #1}}

\newcommand{\Gev}{\text{GeV}}

\newcommand{\beq}{\begin{eqnarray}}
\newcommand{\eeq}{\end{eqnarray}}
\newcommand{\nn}{\nonumber}
\newcommand{\eql}[1]{\label{eq:#1}}
\newcommand{\eq}[1]{(\ref{eq:#1})}

\newcommand{\del}{\partial}
\newcommand{\wbar}[1]{\overline{#1}}

\newcommand{\SU}{\text{SU}}
\newcommand{\U}{\text{U}}
\newcommand{\varep}{\varepsilon}
\newcommand{\hc}{\hbox{h.c.}}
\newcommand{\prpg}[1]{\bigl\langle #1 \bigr\rangle}

\preprint{UMD-PP-08-016}

\title{Supersymmetry in Slow Motion}

\author{Z.~Chacko\\Center for Fundamental Physics, Department of Physics, University of Maryland, College Park, MD 20742, USA\\Department of Physics, University of Arizona, Tucson, AZ 85721, USA}
\author{Christopher A.~Krenke\\Center for Fundamental Physics, Department of Physics, University of Maryland, College Park, MD 20742, USA\\Department of Physics, University of Arizona, Tucson, AZ 85721, USA}
\author{Takemichi Okui\\Center for Fundamental Physics, Department of Physics, University of Maryland, College Park, MD 20742, USA\\Department of Physics and Astronomy, Johns Hopkins University, Baltimore, MD 21218, USA}

\abstract{
We construct new theories of electroweak symmetry breaking that employ a
combination of supersymmetry and discrete symmetries to stabilize the
weak scale up to and beyond the energies probed by the LHC. These models
exhibit conventional supersymmetric spectra but the
fermion-sfermion-gaugino vertices are absent. This closes many
conventional decay channels, thereby allowing several superpartners to
be stable on collider time scales. This opens the door to the
possibility of directly observing R-hadrons and three flavors of
sleptons inside the LHC detectors.}


\begin{document}


\section{Introduction}
\label{sec:intro}

The stability of the weak scale against radiative corrections from
higher scales is a mystery.  While the puzzle itself is conceptual,
virtually all proposed solutions involve new particles and interactions
at the weak scale, providing rich collider phenomenology to be explored
at the Large Hadron Collider (LHC). Since the reach of the LHC is
limited to about 5 TeV, it is important to explore theories of new
physics which can stabilize the weak scale up to these energies, and
lead to interesting collider signals. Several interesting mechanisms
have recently been proposed, including little Higgs
theories~\cite{Little1},~\cite{Little2}, twin Higgs
theories~\cite{twin},~\cite{twinLR}, and folded
supersymmetry~\cite{folded}.

In this paper we propose a new class of models that stabilize the weak
scale up to and beyond the energies probed by the LHC, and which give
rise to exotic collider signatures. These theories are based on the
following observation. In conventional supersymmetric theories such as
the Minimal Supersymmetric Standard Model (MSSM), the quadratically
divergent contributions to the squared Higgs mass in the Standard Model
(SM) are cancelled by the superpartners. However, while this
cancellation relies on the couplings of the superpartners to the Higgs,
conventional supersymmetric collider phenomenology actually depends more
critically on the {\it fermion-sfermion-gaugino} couplings, through
which most superpartners decay. Since the latter interactions do not
play a role in stabilizing the Higgs mass at one loop, the relationship
between supersymmetric naturalness and supersymmetric phenomenology at
LHC energies is somewhat indirect.

This observation then begs the following question. Do there exist
consistent effective field theories that exhibit supersymmetric spectra
and stabilize the weak scale up to the energies probed by the LHC, but
where the fermion-sfermion-gaugino vertices are absent? Such theories
could in general give rise to collider signatures that are completely
different from those of conventional supersymmetric models!

In this paper we construct one realization of such a scenario where
supersymmetry, in combination with a set of discrete symmetries,
stabilizes the weak scale even in the absence of the
fermion-sfermion-gaugino vertices. These discrete symmetries lead to
robust phenomenological consequences. In particular, several
superpartners become stable or quasi-stable on collider time scales,
their decays effectively happening in slow-motion. This opens the door
to the exciting possibility of directly observing several of the
superpartners in the LHC detectors.

This particular realization borrows from the ideas of folded
supersymmetry~\cite{folded}, but the phenomenology of that
case~\cite{BCGHK} (see also \cite{markus}) is very different. For every
quark or lepton superfield in the MSSM, consider adding to the theory an
additional chiral superfield with exactly the same gauge quantum
numbers. We give the MSSM fields a subscript $A$ to distinguish them
from these new fields to which we give a subscript $B$. We impose the
following three $Z_2$ symmetries: $Z_2^{AB}$ which interchanges the $A$
and $B$ fields, $Z_2^A$ and $Z_2^B$ which flip the sign of the $A$ and
$B$ fields, respectively. Then the up-type Yukawa couplings in this
theory take the form
\beq
  \int\! d^2\theta \, y_u (H_u Q_A U_A + H_u Q_B U_B)
\eql{desired-Yukawa}
\eeq
where $H_u$ denotes the up-type Higgs, $Q$ the SU(2) doublet quarks and
$U$ the SU(2) singlet anti-quarks. The discrete symmetries have ensured
the equality of the $A$ and $B$ Yukawa couplings and the absence of
mixing between $A$ and $B$. The one-loop quadratic divergences to the
squared Higgs mass from the fermions are cancelled by loops involving
the corresponding superpartners.

This theory also has a $Z_2^F$ symmetry under which all fermions are odd
and all bosons even.  We now construct a {\it new} theory (the
``daughter'' theory) from this theory by projecting out all states that
are odd under the combined $Z_2^A \times Z_2^F$ symmetry.  Then the
daughter theory only contains the SM fermions from $A$, the scalars from
$B$, and the gauge bosons.  The scalars from $A$, the fermions from $B$
and all the gauginos have been projected out.  We refer to the scalars
from $B$ as the {\it pseudo-sfermions} of the SM fermions from $A$.  The
pseudo-sfermions have the exactly same quantum numbers as the true
sfermions, but do not form supermultiplets with the SM fermions. In
particular, the pseudo-sfermions do not have a vertex with the SM
fermion and the gaugino. Therefore, the daughter theory is fundamentally
non-supersymmetric. Furthermore, $Z_2^{AB}$ is also broken.  However,
provided that the form of \eq{desired-Yukawa} can be justified, the
one-loop quadratic divergences to the squared Higgs mass from the Yukawa
couplings are still cancelled, but now between the SM fermions in $Q_A$
and $U_A$ and the pseudo-sfermions in $Q_B$ and $U_B$.

Of course the daughter theory by itself does not possess any symmetry
that can ensure the equality of the $A$ and $B$ Yukawa couplings
necessary for these cancellations to go through.  Therefore it must
emerge as the low energy limit of some other theory where the equality
of these couplings is a consequence of a symmetry. Furthermore, since
gauginos are required to cancel quadratic divergences from gauge loops,
they must be reintroduced at some level, but without reintroducing
fermion-pseudo-sfermion-gaugino vertices. In the next section we shall
give an explicit example of such a construction.

Remarkably, some phenomenological aspects of this scenario are already
clear. The theory possesses a $Z_2^B$ parity symmetry under which all
the pseudo-sfermions are odd. Therefore the lightest pseudo-sfermion is
stable. Furthermore, notice that the $B$ sector has its own conserved
baryon number, and also three conserved lepton numbers (neglecting the
neutrino masses). Therefore, the lightest pseudo-sfermion with any of
these quantum numbers is necessarily stable. In order to avoid conflict
with the observational bounds on stable charged
particles~\cite{bounds1},~\cite{bounds2},~\cite{bounds3}, these
conservation laws cannot be exact. Since it is technically natural for
the breaking to be small, these pseudo-sfermions can be long-lived or
stable on collider time-scales. Therefore this scenario can give rise to
spectacular signatures involving displaced vertices and stable exotics
at the LHC.

\section{A Model}
\label{sec:model}
Here we present a concrete model which realizes the features presented
above, and constitutes an existence proof of the scenario.  The model
provides a complete self-consistent description up to and beyond LHC
energies, and serves as a useful benchmark for the study of collider
phenomenology.

\subsection{Fields and Symmetries in the Bulk}
\label{sec:bulk}
In this subsection, we focus on the physics in the bulk.  The fields and
symmetries at the boundaries will be discussed in section
\ref{sec:branes}. We take the bulk to be a 5D Minkowski space with the
5th coordinate $x^5 \equiv y$ restricted to an interval $0 \leq y \leq
\pi R$, where $R^{-1}$ is taken to be in the 5-10 TeV range.  We also
assume that the bulk has 5D ${\cal N}=1$ supersymmetry (SUSY).

All SM fields except the Higgs live in the bulk.  (The Higgs will be
located on the boundary at $y=0$.)  The 5D gauge supermultiplets are
denoted by $(A_{aM}, \lambda_a, \lambda_a^c, \sigma_a)$, where $a=1,2,3$
refer to $\U(1)_{\rm Y}$, $\SU(2)_{\rm L}$ and $\SU(3)_{\rm C}$
respectively. Here, $A_{M}$ corresponds to the SM gauge field, $\lambda$
and $\lambda^c$ to the gauginos, and $\sigma$ to the adjoint scalar
which completes the 5D gauge supermultiplet. In accordance with the
scenario outlined above, the matter fields are doubled and we label them
as $q_{ip}$, $u_{ip}$, $d_{ip}$, $\ell_{ip}$ and $e_{ip}$ (all being
left-handed Weyl spinors), where $p$ runs over $A$ and $B$, and
$i=1,2,3$ refer to the three generations. They form supersymmetric
hypermultiplets with the fermions $q^c_{ip}$, $u^c_{ip}$, $\cdots$ in
the corresponding conjugate representations and the scalar partners
$\tilde{q}_{ip}$, $\tilde{u}_{ip}$, $\cdots$, $\tilde{q}^c_{ip}$,
$\tilde{u}^c_{ip}$, $\cdots$.  These fields are collectively referred to
as $\psi_{ip}$, $\psi^c_{ip}$, $\phi_{ip}$, $\phi^c_{ip}$, respectively. 
By definition the SM fermions are the zero modes of $\psi_{iA}$.

The bulk ${\cal N}=1$ SUSY possesses an $\SU(2)_{\rm R}$ symmetry under
which $(\phi_{ip}, \phi_{ip}^{c*})$ and $(\lambda_a, \lambda_a^c)$
transform as doublets.  For our purposes, it is important to distinguish
two different ways of embedding 4D ${\cal N}=1$ multiplets into a 5D
${\cal N}=1$ multiplet. One embedding---the ``unprimed SUSY''---is
$\Phi_{ip} = (\phi_{ip}, \psi_{ip})$ and $\Phi^c_{ip} = (\phi^c_{ip},
\psi^c_{ip})$, while the other---the ``primed SUSY''---is $\Phi'_{ip} =
(\phi_{ip}^{c*}, \psi_{ip})$ and $\Phi^{c\prime}_{ip} = (-\phi_{ip}^*,
\psi_{ip}^c)$.  Similarly, we can have $V_a = (A_{a\mu}, \lambda_a)$ and
$\Sigma_a = (\sigma_a+iA_{a5}, \lambda^c_a)$, or $V'_a = (A_{a\mu},
\lambda^c_a)$ and $\Sigma'_a = (\sigma_a+iA_{a5}, -\lambda_a)$.  The
$\SU(2)_{\rm R}$ symmetry renders irrelevant whether we use the unprimed
or primed basis in the bulk, but the distinction will be important at
the boundaries.

We require the bulk Lagrangian to possess the following $Z_2$ symmetries.
\begin{align}
  Z_2^A &: \Phi_{iA}   \to -\Phi_{iA}    \,,\quad
           \Phi_{iA}^c \to -\Phi_{iA}^c  \,.\eql{Z_2A}\\
  Z_2^B &: \Phi_{iB}   \to -\Phi_{iB}    \,,\quad 
           \Phi_{iB}^c \to -\Phi_{iB}^c  \,.\eql{Z_2B}\\
  Z_2^{AB} &: 
           \Phi_{iA}   \leftrightarrow \Phi_{iB}    \,,\quad
           \Phi_{iA}^c \leftrightarrow \Phi_{iB}^c  \,.\eql{Z_2AB}\\
  Z_2^{\prime AB} &:
           \left\{\begin{aligned}     
             &\Phi'_{iA} \leftrightarrow  \Phi^{c\prime}_{iB}  \,,\quad\>
             \Phi'_{iB} \leftrightarrow -\Phi^{c\prime}_{iA}  \,,\\ 
             &V'_a       \leftrightarrow -(V'_a)^T             \,,\>
             \Sigma'_a  \leftrightarrow -(\Sigma'_a)^T        \,.
           \end{aligned}\right.  \eql{Z_2ABprime}
\end{align}
Note that $Z_2^{AB}$ and $Z_2^{\prime AB}$ together forbid bulk masses
for the matter fields.

\subsection{Fields and Symmetries at the Boundaries}
\label{sec:branes}
Having described the bulk fields and symmetries, we now introduce the
boundaries.  The boundaries do not preserve all the symmetries of the
bulk; in particular, the bulk SUSY is broken by the following boundary
conditions.  For the gauge fields, we choose
\beq
     A_a         :\> (+,+)  \,,
  && \lambda_a   :\> (+,-)  \,, \nn\\
     \sigma_a    :\> (-,-)  \,,
  && \lambda_a^c :\> (-,+)  \,,
\eeq
where the first (second) $\pm$ refers to the boundary condition at $y=0$
($y=\pi R$).  ``$+$'' means the field is allowed to freely fluctuate at
the boundary, while ``$-$'' means the field is constrained, i.e., not an
independent degree of freedom at the boundary.%
\footnote{If we ignore boundary-localized terms, ``$+$'' and ``$-$''
would reduce to the usual Neumann and Dirichlet boundary conditions,
respectively, or ``even'' and ``odd'' in the orbifold language. But
boundary terms are important as we will see later.}
These boundary conditions only preserve the unprimed (primed) SUSY at
$y=0$ ($y=\pi R$).  This is an example of SUSY breaking by the
Scherk-Schwarz mechanism~\cite{SS},~\cite{PQ},~\cite{BHN}. Furthermore,
notice that only the SM gauge fields have zero modes, precisely in
accord with the low-energy spectrum of the scenario described in
Sec.~\ref{sec:intro}.

For the $A$-type matter fields, we choose 
\beq
  && \phi_{iA}   :\> (+,-)  \,, \quad
     \psi_{iA}   :\> (+,+)  \,, \nn\\
  && \phi_{iA}^c :\> (-,+)  \,, \quad
     \psi_{iA}^c :\> (-,-)  \,,
\eql{A-type-matter}
\eeq
while for the $B$ sector, we choose
\beq
  && \phi_{iB}   :\> (+,+)  \,, \quad
     \psi_{iB}   :\> (+,-)  \,, \nn\\
  && \phi_{iB}^c :\> (-,-)  \,, \quad
     \psi_{iB}^c :\> (-,+)  \,.
\eql{B-type-matter}
\eeq
Note that only $\psi_{iA}$ (the SM fermions) and $\phi_{iB}$ (the
pseudo-sfermions) have zero modes. The $\phi_{iB}$ will acquire mass,
but only radiatively. This again exactly realizes the low-energy
spectrum of our scenario. 

Where should the Higgs be located? Note that, of the four bulk $Z_2$
symmetries \eq{Z_2A}-\eq{Z_2ABprime}, $Z_2^{\prime AB}$ is broken at
$y=0$ while $Z_2^{AB}$ is broken at $y=\pi R$. In order to ensure the
desired form of the Yukawa couplings \eq{desired-Yukawa}, it is crucial
to sequester the Higgs from $Z_2^{AB}$ breaking. Therefore, we must
place the Higgs and the Yukawa couplings at $y=0$. The most general
supersymmetric Yukawa couplings consistent with the unbroken $Z_2^A$,
$Z_2^B$ and $Z_2^{AB}$ are given by
\beq
  \mathcal{W}_{y=0} 
    = \sum_{i,j,p} 
        \Bigl( N_{ij}^{(u)}    y_{ij}^{(u)}    H_u Q_{ip} U_{jp}
              +N_{ij}^{(d)}    y_{ij}^{(d)}    H_d Q_{ip} D_{jp}  
              +N_{ij}^{(\ell)} y_{ij}^{(\ell)} H_d L_{ip} E_{jp}
        \Bigr)  \,,
\eql{Yukawas}
\eeq
where $y^{(u,d,\ell)}_{ij}$ are the SM Yukawa couplings, and
$N_{ij}^{(u,d,\ell)}$ account for the normalizations of the zero-mode
wavefunctions:
\beq
  N_{ij}^{(u)} 
  \equiv \frac{1}{\xi_{q_{iA}}^{(0)}(0) \, \xi_{u_{jA}}^{(0)}(0)}  
         \quad\text{etc.,} \eql{normalizations} 
\eeq 
where $\xi_{q_{iA}}^{(n)} (y)$ is the normalized $n$-th KK mode of
$q_{iA}$, etc. (We have also implicitly imposed an R parity, which has
the same charge assignments as in the MSSM for each of the fields in $A$
and $B$.)

The symmetries at $y=0$ also allow brane-localized kinetic terms at this
point. The $Z_2^{AB}$ symmetry ensures that the kinetic terms for the
$A$ and $B$ fields have the same coefficient, and therefore the
cancellation of the quadratic divergences arising from Yukawa couplings
is maintained. On the other hand, at the $y=\pi R$ boundary, the only
quadratic terms allowed by $Z_2^A$, $Z_2^B$ and $Z_2^{\prime AB}$ are
the brane-localized kinetic terms for $\Phi'_{iA}$ and
$\Phi_{iB}^{c\prime}$.  $Z_2^{\prime AB}$ ensures that these two kinetic
terms have the same coefficients, which again guarantees the
cancellation.

In summary, our choice of boundary conditions ensures that the only zero
modes arising from the bulk fields are the SM fermions from the
$A$-sector, the pseudo-sfermions from the $B$-sector, and the SM gauge
bosons.  The discrete symmetries relating the $A$ and $B$ fields ensure
that one loop quadratic divergences to the Higgs mass arising from loops
involving the SM fermions are cancelled by the pseudo-sfermions. What
about loops involving the gauge fields? The lightest gauginos have
masses of order $1/2R$. The fact that the above boundary conditions
break supersymmetry only non-locally ensures that contributions to the
Higgs mass from gauge loops are finite and cutoff at this scale.
Provided that the compactification scale $1/R$ is less than about 5-10
TeV, radiative corrections to the Higgs mass from gauge loops are under
control even though there is no fermion-pseudo-sfermion-gaugino vertex. 
Therefore this model is a concrete realization of the scenario described
in Sec.~\ref{sec:intro}.

\subsection{The Long-lived Pseudo-sfermions}

Now, as anticipated in the introduction, this model as it stands
conserves $B$-baryon and $B$-lepton numbers, implying that the lightest
$B$ scalars are necessarily stable.  This is incompatible with
observation. To resolve this problem, we allow for small violations of
$Z_2^A$ and $Z_2^B$ by adding to \eq{Yukawas} the following terms:
\beq
  \Delta{\mathcal{W}}_{y=0}
    &=& \sum_{i,j} 
        \Bigl( N_{ij}^{(u)} y_{ij}^{(u)} \epsilon_{ij}^{(u)} H_u Q_{iA} U_{jB}
              +N_{ij}^{(d)} y_{ij}^{(d)} \epsilon_{ij}^{(d)} H_d Q_{iA} D_{jB}              
              +N_{ij}^{(\ell)} y_{ij}^{(\ell)} \epsilon_{ij}^{(\ell)} 
               H_d L_{iA} E_{jB}  
        \Bigr) \nn\\
    & & +(A \leftrightarrow B)  \,,\eql{AB-mixing}
\eeq
where $\epsilon_{ij}^{(u,d,\ell)}$ are dimensionless parameters, and
$N_{ij}^{(u,d,\ell)}$ are given in \eq{normalizations}. Note that this
still preserves $Z_2^{AB}$, so the cancellation of the quadratic
divergences in the squared Higgs mass is not spoiled. 

One technical remark is in order. Note that while there are no terms
with a lower dimension than those in \eq{AB-mixing} which can mix $A$
and $B$ while preserving $Z_2^{AB}$, brane-localized kinetic mixing
terms, such as $Q_{iA}^\dag e^{2V} Q_{jB}$, have the same dimension.
However, unlike the mixed Yukawa couplings above, such terms do not
affect the zero modes and therefore do not give rise to decays of the
light scalars.  If we begin with such kinetic $A$-$B$ mixing but without
the mixed Yukawas \eq{AB-mixing}, then the mixed Yukawas would never be
induced, due to the non-renormalization theorem of superpotential. On
the other hand, if we begin with mixed Yukawas but without kinetic
$A$-$B$ mixing, it would be induced radiatively.  Since the only
phenomenological relevance of the $A$-$B$ mixings is to cause
$\phi_{iB}^{(0)}$ to decay, we neglect the brane-localized kinetic
$A$-$B$ mixings, which can self-consistently be assumed to be suppressed
by a loop factor with respect to the mixed Yukawa couplings.

\subsection{The Cutoff of the Theory}
\label{sec:cutoff}

Here we estimate the scales suppressing higher dimensional operators in
our Lagrangian, which we have neglected in our analysis up till now. In
principle, locality allows three separate scales, i.e.~the cutoffs at
$y=0$, in the bulk, and at $y=\pi R$, which we denote by $\Lambda_0$,
$\Lambda_b$, and $\Lambda_{\pi R}$, respectively. We take the true
cutoff of the model to be the lowest of the three.

\FIGURE[t]{\epsfig{file=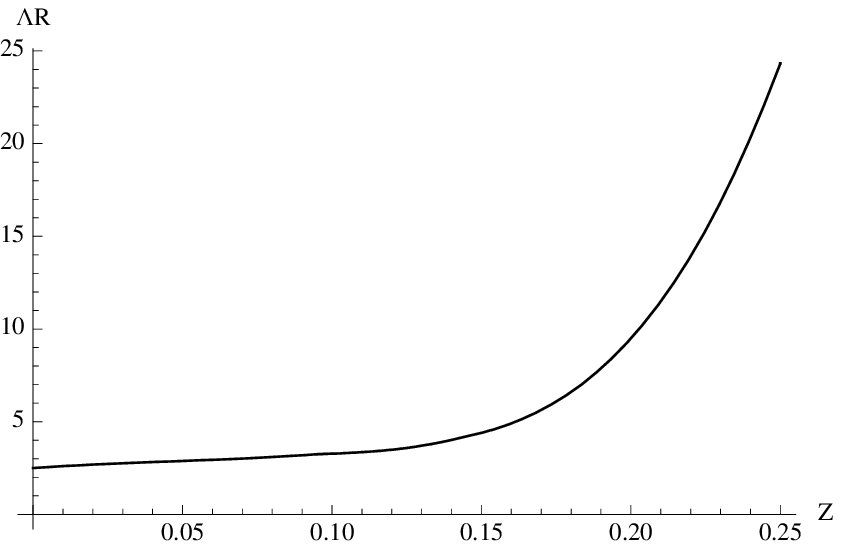, width=5in}\caption{The cutoff $\Lambda$ of the model as a function of $Z$.}\label{fig:cutoff}}

The most severely constrained is $\Lambda_0$, because the
brane-localized top Yukawa interactions can rapidly become strong above
$1/R$ due to the $O(1)$ top Yukawa coupling, color multiplicity, and the
fact that the Higgs couples to both $A$ and $B$ fields. Since there is
no large or small number involved here, one might naively expect that
$\Lambda_0 \sim 1/R$, i.e.~no significant gap between the
compactification scale and the cutoff. Then the effects of higher
dimensional operators can potentially be large, invalidating the 5D
effective field theory.
 
However, in estimating $\Lambda_0$, it is crucial {\it not} to neglect
the brane-localized kinetic terms, because their effects are larger for
the heavier KK modes which are important when analyzing the UV behavior.
In terms of the dimensionless coefficient $Z$ defined via
\beq
  {\cal L}_\text{brane kinetic} 
  = \int\! d^4\theta \, Z \pi R \, \Phi^\dag e^{2V} \Phi \Bigr|_{y=0}  \,,
\eql{brane-Z}
\eeq
a rough estimate of the cutoff $\Lambda$ ($=\Lambda_0$) as a function of
$Z$ (taken to be equal for $Q_3$ and $U_3$ for simplicity) is plotted in
Fig.~\ref{fig:cutoff}. As one can see, the cutoff $\Lambda_0$ can easily
exceed $\sim O(10)/R$ for $Z\gsim \text{a few}/10$. This then becomes
comparable to the bound on $\Lambda_b$ from gauge loops in the bulk, and
there is no advantage to further raising $Z$. 

The final check is to make sure that such values of $Z$ do not
significantly lower the masses of the lightest KK fermions in the $B$
sector, which would jeopardize our scenario. The lightest KK $B$-fermion
mass can be computed from \eq{KKmasses} in Appendix \ref{app:KKmodes},
and one finds that when varying $Z$ from 0 to 0.25 the mass only
decreases from $0.5/R$ to $0.4/R$. Therefore, we conclude that our 5D
model with $R^{-1}$ in the 5-10 TeV range provides a good effective
field theory realization of our scenario up to and beyond energies
accessible to the LHC.

One might wonder if the overall cutoff of the theory could be raised
simply by warping the 5D spacetime. This possibility is however
difficult to reconcile with the Scherk-Schwarz mechanism for
supersymmetry breaking~\cite{Take}, which plays a crucial role in our
construction.

\subsection{The Pseudo-sfermion Spectrum}
\label{sec:spectrum}

The one-loop gauge contributions to the squared soft mass of the scalar
$i$ are given by
\beq
  \delta m^2_{i,\text{gauge}}
  &=& \frac{1}{4\pi^4 R^2} \frac{K_g(Z_i)}{1+Z_i} \sum_G g_G^2 C^{(G)}_2(i)  \,,
\eql{softmass-gauge}
\eeq
where $G=\U(1)_Y, \SU(2)_{\rm L}, \SU(3)_{\rm C}$, and $g_G$ and
$C_2^{(G)}$ are respectively the gauge coupling and the quadratic
Casimir for the group $G$.  $K_g(Z)$ is an $O(1)$ dimensionless
integral:
\beq
  K_g(Z) 
  \equiv \int_0^\infty\!\!\! dx
         \frac{x^2 \!\Bigl[ 2-Z + \frac{Z^3 x^2}{2} 
               +Z(1 + 2 Z) x \coth x \Bigr]}
              {(4 + Z^2 x^2)\sinh x + 4Zx \cosh x} .\!\!\!\!\nn
\eql{Kg(Z)}
\eeq
The combination $K_g(Z)/(1+Z)$ is represented in Fig.~\ref{fig:KgKY} by
the solid curve.

\FIGURE[t]{\epsfig{file=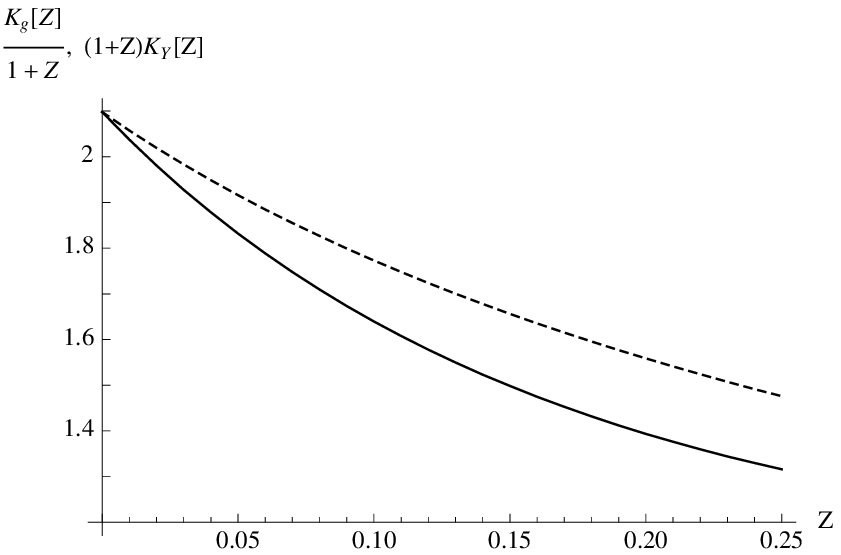, width=5in}\caption{The functions $K_g(Z)/(1+Z)$ (solid curve) and $(1+Z)\, K_Y(Z)$ (dashed curve).}\label{fig:KgKY}}

For the third generation squarks, the one-loop contributions from the
large top Yukawa coupling are also important. 
\beq
  \delta m_{\tilde{q}_3}^2
  &=& \frac{y_t^2}{8\pi^4 R^2} (1+Z_{U_3}) \, K_Y(Z_{U_3})  \,,\nn\\
  \delta m_{\tilde{u}_3}^2
  &=& \frac{y_t^2}{4\pi^4 R^2} (1+Z_{Q_3}) \, K_Y(Z_{Q_3})  \,,
\eql{softmass-Yukawa}
\eeq
where
\beq
  K_Y(Z) 
  = \int_0^\infty\!\! dx\, \frac{2x^2}{(4+Z^2 x^2)\sinh x + 4Zx\cosh x}  \,.
\eql{KY(Z)}
\eeq
The combination $(1+Z)\, K_Y(Z)$ is represented in Fig.~\ref{fig:KgKY}
by the dashed curve.

\subsection{The Higgs Sector and Electroweak Symmetry Breaking}

There are a few points regarding the Higgs sector and electroweak
symmetry breaking which must be discussed.
First, recall that the bulk possesses an
$\SU(2)_{\rm R}$ symmetry, which contains a $\U(1)_{\rm R}$ subgroup
generated by rotations about the $\sigma^3$ direction.  In the
convention where the gauginos $\lambda$ and $\lambda^c$ have $\U(1)_{\rm
R}$ charges $+1$ and $-1$, $\Phi_{ip}$ and $\Phi^c_{ip}$ have
$\U(1)_{\rm R}$ charges $+1$ and $-1$, respectively, while $V_a$ and
$\Sigma_a$ have no $\U(1)_{\rm R}$ charge.  The bulk also possesses a
global $\U(1)$ symmetry under which $\Phi_{ip}$ and $\Phi^c_{ip}$ have
charge $-1/2$ and $+1/2$ respectively.  The sum of this $\U(1)$ and the
above $\U(1)_{\rm R}$ is also an R-symmetry, which we label $\U(1)_{\rm
R'}$.  Then, the superpotential \eq{Yukawas} is $\U(1)_{\rm R'}$
invariant if $H_u$ and $H_d$ are assigned unit $\U(1)_{\rm R'}$ charge. 
The $\mu$-term $\delta(y)\int\! d^2 \theta \; \mu H_u H_d$ also respects
$\U(1)_{\rm R'}$.  Therefore, as things stand, the theory is $\U(1)_{\rm
R'}$ invariant, and a $B\mu$-term $B\mu \tilde{H}_u \tilde{H}_d$ will
not be generated.

Therefore, in order to have realistic electroweak symmetry breaking, we
must introduce explicit $\U(1)_{\rm R'}$ violation.  It is also
desirable to have additional contributions to the Higgs quartic
couplings to get a Higgs heavier than the LEP bound without too much
tuning.  As shown in {\cite{folded}}, both objectives can be
simultaneously realized by introducing additional SM singlets in the Higgs
sector.  Below, we summarize this analysis.

We extend the Higgs sector by adding to the theory an extra singlet $S$ which is localized to the 
brane at $y=0$ and replaces the $\mu$ term 
$\delta(y)\int\! d^2 \theta \; \mu H_u H_d$ by 
\beq
\delta \left( y \right) \int d^2 \theta \left[\alpha S + \lambda S H_u H_d + \kappa 
S^3 \right]  \,.
\eeq
Now $\U(1)_{\rm R'}$ is explicitly broken.
The Higgs sector has no continuous global symmetries, which ensures 
the absence of an unwanted Goldstone boson.
The Higgsino mass, or the $\mu$ term, will be supplied by the VEV of the scalar
$S$.  (The origin of the negative squared mass for the scalar $S$ will be 
discussed below.)
The above superpotential also provides an additional tree-level Higgs quartic
coupling.  For example, for $\tan\beta \sim O(1)$ and $\lambda \gsim 0.7$, 
the tree-level Higgs masses will be greater than the experimental lower bound.
Such ``large'' values of $\lambda$ are allowed since the cutoff of the theory
is low.
We choose the value of $\alpha$ to be of order weak scale size to obtain consistent electroweak breaking. This choice is
technically natural.  We leave the problem of naturally generating $\alpha$ of this size
for future work.

Now we are ready to compute the radiatively generated soft mass of the Higgs.
First, there is a one-loop contribution from gauge loops, given by the formula \eq{softmass-gauge}:
\beq
  \delta m_H^2|_{\rm gauge}
  = \frac{2.1}{4\pi^4 R^2} \left( \frac{3g_2^2}{4} + \frac{g_1^2}{4} \right)
  \simeq \frac{0.075}{4\pi^2} R^{-2}  \,, 
\eeq
where we have ignored $Z_H$. (The values corresponding to nonzero $Z_H$ can be read off from 
Fig.~\ref{fig:KgKY}.)
There is a two-loop contribution from top and stop loops where the stop masses are generated at one loop
given by the formulae \eq{Kg(Z)} and \eq{KY(Z)}, giving rise to
\beq
  \delta m_H^2|_{\rm top}
  \simeq -\frac{3y_t^2}{4\pi^2} \widetilde{m}_t^2 \log\frac{R^{-1}}{\widetilde{m}_t}  \,,
\eeq
where $\widetilde{m}_t^2$ is the average of the left and right stop mass-squareds. 
Taking $Z_{Q_3} = Z_{U_3} \equiv Z_t$ 
for simplicity, one finds that $\widetilde{m}_t^2$ varies from $0.017 R^{-2}$
to $0.012 R^{-2}$ as $Z_t$ is varied from 0 to 2. In this range the total $\delta m_H^2$ 
is negative, thereby triggering electroweak symmetry breaking. 

Finally, let us discuss the origin of a negative squared mass for the scalar in $S$.
A simple way to generate this is to introduce into the bulk two SM singlet hypermultiplets $\hat{P}_A$ 
and $\hat{P}_B$. The boundary conditions on these fields are such as to allow only a fermion zero mode
for each of $\hat{P}_A$ and $\hat{P}_B$. The bulk $Z_{AB}$ symmetry interchanges $\hat{P}_A$ and
$\hat{{P}}_B$. In addition, under the $Z_{AB}'$ symmetry, $\hat{P}_A$ and $\hat{{P}}_B$ are 
also interchanged.  
Then on the brane at $y=0$ we can write the interaction
\begin{equation}
\delta \left( y \right) \int d^2 \theta \left[ \lambda_P S P_A P_B + 
\mu_{P} \left({P_A}^2 + {P_B}^2 \right) \right]
\end{equation}
The effect of the coupling $\lambda_P$ is to generate a negative mass squared for the scalar in
$S$ at one loop.  
Note that the theory possesses a $Z_2$ symmetry under which $P_A \to -P_A$ and $P_B \to -P_B$ while
all other fields are invariant.  Then, the lightest fermion in $P_{A,B}$ is stable and therefore can
be a viable dark matter candidate \cite{pedestrian}.

The extended Higgs sector has no impact on the essential aspects of the theory such as
the mechanism for cancellation of the one-loop quadratic divergences, or 
the lifetimes of the long-lived pseudo-sfermions.  
Further, the new couplings introduced above
do not affect our estimate of the cutoff of the theory in section \ref{sec:cutoff}.
The reason is that this estimate is dominated by the behavior of the top Yukawa coupling, 
which rapidly grows strong above $1/R$, whereas the new couplings in the Higgs 
sector need not run rapidly.

\section{Collider Phenomenology}
\label{sec:pheno}

Let us first contrast the phenomenology of this scenario with that of
the MSSM (and extensions of the MSSM that include additional singlets)
with a similar spectrum.  We therefore consider a spectrum where the
gauginos are heavy, at a few TeV, while the sfermions and Higgsinos are
at a few to several hundred GeV.  We further specialize the case where
the Higgsino is the LSP. In the MSSM with such a spectrum, gauginos are
not directly accessible to the LHC. Therefore, sfermions predominantly
decay to the corresponding SM fermions and a Higgsino. These decays are
{\it prompt.}

In stark contrast, in our scenario, the lightest pseudo-sfermions
(i.e.~the lightest pseudo-squark and the three lightest $e$-, $\mu$-,
$\tau$-type pseudo-sleptons) can decay only via the couplings
\eq{AB-mixing}. Since these couplings break symmetries, it is
technically natural for them to be small. Therefore, these four lightest
sfermions can be {\it naturally long-lived or even collider-stable!}. 

In our specific extra-dimensional construction, the form of the soft
masses \eq{softmass-gauge} implies that SU(2) doublet pseudo-sfermions
are heavier than the SU(2) singlet ones with the same baryon, $e$-,
$\mu$-, or $\tau$-number.  Thus, the doublets will promptly decay to the
corresponding singlet scalars. Neglecting the masses of the decay
products, the SU(2) singlet pseudo-sfermions' decay rates are given by
\beq
  \Gamma^{-1} 
    = \biggl( \frac{y^2\varep^2}{8\pi} \widetilde{m} \biggr)^{\!-1}
    \simeq 50 \>{\rm \mu m}\> \frac{100\>\Gev}{\widetilde{m}}
           \biggl( \frac{10^{-6}}{y\varep} \biggr)^{\!2}  \,,
\eeq
where $\widetilde{m}$ is the pseudo-sfermion mass, while $y\varep$
represents the relevant combination of the Yukawa coupling and the
$\varep$ factors as in \eq{AB-mixing}.  For example, the
pseudo-selectron could have a displaced vertex for $\varep \sim O(0.1)$
and $\tan\beta \sim O(1)$.  It is also possible that two pseudo-sleptons
of different generations both decay inside the detector.  For example,
for $\varep \sim O(10^{-3})$ and $\tan\beta \sim O(1)$, a pseudo-smuon
will have a displaced vertex of about 100 $\mu m$, while a
pseudo-selectron decays after travelling a meter or so.

Due to their large masses, these long-lived charged pseudo-sleptons
hardly lose any energy while coasting through the detecter
material~\cite{DX}. If the produced pair of the pseudo-sleptons are
collider-stable, we expect to see two highly-ionizing tracks. On the
other hand, if each decays into a SM fermion and a Higgsino in the
detector, there will be two tracks with a kink. In this regard, our
scenario shares some similarity with the slepton co-NLSP
scenario~\cite{DDRT} in gauge mediation.

The long-lived pseudo-squarks, on the other hand, will hadronize into
$R$-hadrons, which may be neutral or charged.  If charged, they will
again appear as a highly-ionizing tracks, although a neutral $R$-hadron
can sometimes be converted to a charged one by interacting with nucleons
in the detector.  Slow enough $R$-hadrons can be stopped~\cite{ADPRW},
as in the case of the long-lived gluino in Split
Supersymmetry~\cite{splitSUSY}.
   
Note that these signals are quite robust expectations of our low-energy
scenario described in Sec.~\ref{sec:intro}, independent of the details
of any particular UV completion, and should make this scenario
straightforward to distinguish at the LHC.

\begin{acknowledgments}

We thank Roni Harnik for collaboration during the early stages of this
work. T.O.~is supported in part by DOE grant DE-FG02-03ER4127 and by the
Alfred P.~Sloan Foundation. Z.C.~and C.A.K.~are supported by the NSF
under grant PHY-0801323.  We also thank Aspen Center for Physics where part
of this work was conducted.

\end{acknowledgments}

\appendix   

\section{Boundary Conditions with Boundary-Localized Kinetic Terms}
\label{app:bc}
We analyze a massless bulk fermion with a brane-localized kinetic term
at the $y=0$ boundary.  First, consider the contribution to the 4D
action from the 5D {\it bulk} kinetic term:
\beq
  {\cal L}_{\rm 4D}
  \equiv \int_0^{\pi R}\!\! dy\, {\cal L}_\text{bulk kin.}  
         \quad \text{(Naive),}
\eql{naive}
\eeq
where
\beq
  {\cal L}_\text{bulk kin.}
  = \wbar{\psi}\, p\cdot\bar{\sigma} \psi + \psi^c p\cdot\sigma \wbar{\psi}^c
   +\frac12 \left[ \psi^c \del_y \psi - (\del_y \psi^c) \psi +\hc \right]  \,.
\eeq
(We have carefully distributed $\del_y$ such that ${\cal L}_\text{bulk
kin.}$ is real without integration by parts.) The reason that \eq{naive}
is ``naive'' is the following.  From the equations of motion in the bulk
\beq
  p\cdot\bar{\sigma} \psi - \del_y \wbar{\psi}^c &=& 0  \,,\nn\\
  p\cdot\sigma \wbar{\psi}^c + \del_y \psi       &=& 0  \,,
\eql{eom}
\eeq
we expect that ${\cal L}_{\rm 4D}$ can depend on only two of the four
variables $\psi(0)$, $\psi^c(0)$, $\psi(\pi R)$, and $\psi^c(\pi R)$. 
For example, when $\psi$ and $\psi^c$ have the $A$-type boundary
conditions ($(+,+)$ and $(-,-)$ respectively), ${\cal L}_{\rm 4D}$ by
definition should only depend on $\psi(0)$ and $\psi(\pi R)$.  Thus,
$\delta {\cal L}_{\rm 4D}$ should only contain $\delta\psi(0)$ and
$\delta\psi(\pi R)$.  Similarly, when $\psi$ and $\psi^c$ have the
$B$-type boundary conditions ($(+,-)$ and $(-,+)$ respectively),
$\delta{\cal L}_{\rm 4D}$ should only depend on $\delta\psi(0)$ and
$\delta\psi^c(\pi R)$. However, notice that the variation of \eq{naive}
is
\beq
  \delta {\cal L}_{\rm 4D}
  = \frac12 \left[ \psi^c \delta \psi - \delta \psi^c \, \psi 
                  +\hc 
            \right] \biggr|_{y=0}^{y=\pi R} 
\eql{wrong}
\eeq
which depends on all of $\delta\psi(0)$, $\delta\psi^c(0)$,
$\delta\psi(\pi R)$, and $\delta\psi^c(\pi R)$.
  
We usually fix this problem by hand by imposing the boundary condition
$\psi^c(0)=\psi^c(\pi R)=0$ for the $A$-type, or $\psi^c(0)=\psi(\pi
R)=0$ for the $B$-type.  This is analogous to imposing constraints {\it
by hand} when solving a constrained mechanical system.  Alternatively,
we can let {\it mathematics} take care of the constraints by adding
Lagrange multipliers. In our case, a suitable mathematical trick is to
add to \eq{naive} the following terms: for the $A$-type
\beq
  {\cal L}_{\rm 4D}^{(A)}
  \equiv  \int_0^{\pi R}\!\! dy\, {\cal L}_\text{bulk kin.}  
         +\frac12\left[ -\psi^c\psi(0) + \psi^c\psi(\pi R) +\hc \right]  
          \quad \text{(Correct)}  \,,\nn
\eql{correct}
\eeq
while for the $B$-type
\beq
  {\cal L}_{\rm 4D}^{(B)}
  \equiv  \int_0^{\pi R}\!\! dy\, {\cal L}_\text{bulk kin.}  
         +\frac12\left[ -\psi^c\psi(0) - \psi^c\psi(\pi R) +\hc \right]
          \quad \text{(Correct)} \nn
\eeq
Then, instead of \eq{wrong}, we now obtain
\beq
  \delta {\cal L}_{\rm 4D}^{(A)}
  = -\psi^c \delta \psi (0) + \psi^c \delta \psi (\pi R) + \hc  \,,
\eql{A-variation}
\eeq
and
\beq
  \delta {\cal L}_{\rm 4D}^{(B)}
  = -\psi^c \delta \psi (0) - \delta \psi^c \psi (\pi R) + \hc  \,.
\eql{B-variation}
\eeq
Note that \eq{A-variation} shows that ${\cal L}_{\rm 4D}^{(A)}$ is a
function of only $\psi(0)$ and $\psi(\pi R)$ as it should be. In the
absence of other terms at the boundaries, demanding that $\delta {\cal
L}_{\rm 4D}^{(A)}$ vanish for arbitrary variations gives us the usual
boundary conditions $\psi^c(0)=\psi^c(\pi R)=0$, which, together with
the equations of motion \eq{eom}, further implies that $\del_y\psi(0) =
\del_y\psi(\pi R) = 0$.  This is what we would have got in the orbifold
language by assigning $(+,+)$ and $(-,-)$ parities to $\psi$ and
$\psi^c$ respectively.  Similarly, demanding that the variation
\eq{B-variation} vanish is equivalent to assigning $(+,-)$ and $(-,+)$
parities to $\psi$ and $\psi^c$.

The advantage of using ${\cal L}_{\rm 4D}^{(A)}$ and ${\cal L}_{\rm
4D}^{(B)}$ becomes clear once there are extra terms at the boundaries. 
For example, consider the $A$-type case and let us add a
boundary-localized kinetic term at $y=0$:
\beq
  {\cal L}_Z^{(A)}
   = {\cal L}_{\rm 4D}^{(A)} 
    +Z\pi R \,\wbar{\psi}\, p\cdot\bar{\sigma} \psi(0)  \,.
\eeq
Then, extremizing ${\cal L}_Z^{(A)}$ readily give us
\beq
  \wbar{\psi}^c (0) &=& Z\pi R \, p\cdot\bar{\sigma} \psi(0)  \,,\nn\\
  \psi^c(\pi R)     &=& 0  \,.
\eeq
Combining these with the equations of motion \eq{eom}, we obtain
\beq
  \del_y\psi(0)     &=& -Z\pi R\, p^2 \psi(0)  \,,\nn\\
  \del_y\psi(\pi R) &=& 0  \,,
\eql{++}
\eeq
and
\beq
  Z\pi R \, \del_y\psi^c(0) &=& \psi^c (0)  \,,\nn\\
  \psi^c(\pi R)             &=& 0  \,.
\eql{--}
\eeq
These are the correct boundary conditions for the $A$-type fermion with
a brane-localized kinetic term at $y=0$.

Similarly, for the $B$-type fermion, extremizing    
\beq
  {\cal L}_Z^{(B)}
   = {\cal L}_{\rm 4D}^{(B)} 
    +Z\pi R \,\wbar{\psi}\, p\cdot\bar{\sigma} \psi(0)
\eeq
gives
\beq
  \wbar{\psi}^c (0) &=& Z\pi R \, p\cdot\bar{\sigma} \psi(0)  \,,\nn\\
  \psi(\pi R)     &=& 0  \,,
\eeq
which, combined with \eq{eom}, implies
\beq
  \del_y\psi(0) &=& -Z\pi R\, p^2 \psi(0)  \,,\nn\\
  \psi(\pi R)   &=& 0  \,,
\eql{+-}
\eeq
and
\beq
  Z\pi R \, \del_y\psi^c(0) &=& \psi^c (0)  \,,\nn\\
  \del_y \psi^c(\pi R)      &=& 0  \,.
\eql{-+}
\eeq
These are the correct boundary conditions for the $B$ type fermion with
a brane-localized kinetic term at $y=0$.

Of course, one could re-derive all of the above results in the orbifold
language, but must be careful in doing so, because, at the $y=0$ brane,
``odd'' fields jump while even fields have a kink. So, the usual
advantage of the orbifold language, namely the simple relation between
the parity of a field and its boundary condition, is lost.

Although the above analysis was done for fermions, it should be obvious
that, by supersymmetry, bulk scalars with $(+,+)$, $(-,-)$, $(+,-)$, and
$(-,+)$ parities obey the same boundary conditions, \eq{++}, \eq{--},
\eq{+-}, and \eq{-+}, respectively.

\section{The KK Modes}
\label{app:KKmodes}
We expand an $A$-type 5D fermion field as
\beq
  \psi(p,y)   &=& \sum_n \xi_n^{++} (y) \psi_n(p)  \,,\nn\\
  \psi^c(p,y) &=& \sum_n \xi_n^{--} (y) \psi^c_n(p)  \,,
\eeq
where $\Psi_n \equiv (\psi_n, \wbar{\psi}^c_n)$ satisfies the 4D Dirac
equation $\sla{p}\Psi_n = m_n \Psi_n$.  Then, the 5D Dirac equation
\eq{eom} implies that both $\xi(y)$'s satisfy the bulk equation of
motion $(m_n^2 + \del_y^2) \xi_n(y) = 0$ in the interval $0<y<\pi R$.
The boundary conditions for $\xi^{++}_n$ and $\xi^{--}_n$ are given by
\eq{++} and \eq{--}, respectively.  It is trivial to repeat the exercise
for the $B$-type fermion and also for scalars.

The solutions for the $(+,+)$ and $(-,-)$ modes are then given by
\beq
  && \left\{ \begin{array}{ccc}
        \xi^{(++)}_0 (y) &=& 1/\sqrt{\pi R (1+Z)}  \cr
        \xi^{(++)}_n (y) &=&  N(m_n^{\pm\pm}) \,\cos [m_n^{\pm\pm} (y-\pi R)]
     \end{array} \right. \eql{xi++}\\
  && \xi^{(--)}_n (y) = N(m_n^{\pm\pm}) \,\sin [m_n^{\pm\pm} (y-\pi R)]  
       \,,\eql{xi--}
\eeq
where $n=1,2,3,\cdots$, while for the $(+,-)$ and $(-,+)$ modes
\beq
  && \xi^{(+-)}_n (y) =  N(m_n^{\pm\mp}) \,\sin [m_n^{\pm\mp} (y-\pi R)]  
       \,,\eql{xi+-}\\
  && \xi^{(-+)}_n (y) = -N(m_n^{\pm\mp}) \,\cos [m_n^{\pm\mp} (y-\pi R)]  
       \,,\eql{xi-+}
\eeq
where again $n=1,2,3,\cdots$.  The corresponding KK masses and
normalization factors are given by
\beq
  m_0^{++}               &=&  0  \,, \\
  Z \pi R\, m_n^{\pm\pm} &=& -\tan (m_n^{\pm\pm} \pi R)  \,,\\
  Z \pi R\, m_n^{\pm\mp} &=&  \cot (m_n^{\pm\mp} \pi R)  \,,
\eql{KKmasses}
\eeq
and
\beq
  N(m)
  = \sqrt{\frac{2}{\pi R}} 
    \left( 1 + \frac{Z}{1+(Z\pi R \,m)^2} \right)^{-1/2}  .
\eeq

As they should be, all these modes are orthonormal.  Note that, because
of the brane-localized kinetic term, the appropriate inner-products are
\beq
  \braket{f}{g} = Z\pi R \, f^*(0) \, g(0) 
                 +\int_0^{\pi R}\!\! f^*(z) \, g(z) \, dz  \,,
\eeq
for the ``$(+,+)$'' and ``$(+,-)$'' fields, while
\beq
  \braket{f}{g} = \int_0^{\pi R}\!\! f^*(z) \, g(z) \, dz  \,,
\eeq
for the ``$(-,+)$'' and ``$(-,-)$'' types.

\section{The Propagators}

First, consider four massless bulk scalars $\phi_{\alpha\alpha'}$ with
$\alpha, \alpha' = \pm, \pm$. First, all four of them satisfy the bulk
equation of motion $(p^2 + \del_y^2) \phi_{\alpha\alpha'}(p,y) = 0$ in
the interval $0<y<\pi R$, so all the propagators satisfy
\beq
  (p^2 + \del_y^2 + i\epsilon) G_{\alpha\alpha'} (y,y';p) = i\delta(y-y')
\eeq
in this interval.  Then, viewing $G_{\alpha\alpha'} (y,y';p)$ as a
function of $y$, it satisfies \eq{++}, \eq{--}, \eq{+-}, and \eq{-+} for
$(\alpha, \alpha')=(+,+)$, $(-,-)$,$(+,-)$, and $(-,+)$, respectively.
For example, 
\beq
  \del_y G_{++}(y,y';p) \bigr|_{y \to 0}     
  &=& -Z\pi R\, p^2 G_{++}(y,y';p) \bigr|_{y \to 0}  \,,\nn\\
  \del_y G_{++}(y,y';p) \bigr|_{y \to \pi R} 
  &=& 0  \,.
\eeq
Solving these, we obtain
\beq
  G_{++}(y, y', p; Z)
  = -\frac{i\cosh[p_E (y_>-\pi R)] \, 
          \bigl( \cosh[p_E y_<] + Z\pi R \,p_E \sinh[p_E y_<] \bigr)}
         {p_E \bigl(\sinh[p_E\pi R] 
                       +Z\pi R \,p_E \cosh[p_E\pi R] \bigr)}  \,,
\eeq
where $p_E \equiv (-p^2-i\epsilon)^{1/2}$, and $y_>$ and $y_<$ are
respectively the larger and the smaller of $y$ and $y'$.  Similarly, we
have
\beq
  G_{--}(y, y', p; Z)
  &=& \frac{i\sinh[p_E (y_>-\pi R)] \,
             \bigl( \sinh[p_E y_<] + Z\pi R \,p_E \cosh[p_E y_<] \bigr)}
            {p_E \bigl(\sinh[p_E\pi R] 
                          +Z\pi R \,p_E \cosh[p_E\pi R] \bigr)}  \,,\\
  G_{+-}(y, y', p; Z)
  &=& \frac{i\sinh[p_E (y_>-\pi R)] \,
             \bigl( \cosh[p_E y_<] + Z\pi R \,p_E \sinh[p_E y_<] \bigr)}
            {p_E \bigl(\cosh[p_E\pi R] 
                          +Z\pi R \,p_E \sinh[p_E\pi R] \bigr)}  \,,\\
  G_{-+}(y, y', p; Z)
  &=& -\frac{i\cosh[p_E (y_>-\pi R)] \, 
            \bigl( \sinh[p_E y_<] + Z\pi R \,p_E \cosh[p_E y_<] \bigr)}
           {p_E \bigl(\cosh[p_E\pi R] 
                         +Z\pi R \,p_E \sinh[p_E\pi R] \bigr)}  \,.
\eeq

Using these scalar propagators, we can also write down the propagators
for fermions. For the $A$-type fermion, we have chirality-preserving
propagators
\beq
  \prpg{\psi_\alpha (y) \, \wbar{\psi}_{\dot{\beta}} (y')}(p)
     &=& p \!\cdot\! \sigma_{\alpha \dot{\beta}} \, G_{++}(y,y',p;Z)  \,,\nn\\
  \prpg{\wbar{\psi}^{c\dot{\alpha}} (y) \, \psi^{c\beta} (y')}(p)
     &=& p \!\cdot\! \bar{\sigma}^{\dot{\alpha} \beta} \, G_{--}(y,y',p;Z)  \,,
\eeq
and chirality-flipping propagators
\beq
  \prpg{\psi_\alpha (y) \psi^{c\beta} (y')}(p)
     &=& \delta_\alpha^\beta \, \del_y G_{--}(y,y',p;Z)  \,,\nn\\
  \prpg{\wbar{\psi}^{c\dot{\alpha}} (y) \, \wbar{\psi}_{\dot{\beta}} (y')}(p)
     &=& -\delta^{\dot{\alpha}}_{\dot{\beta}} \, \del_y G_{++}(y,y',p;Z)  \,,
\eeq
where $\langle\cdots\rangle(p)$ denotes the time-ordered correlation
function in the mixed momentum-position representation. Similarly, for
the $B$-type fermion, we have
\beq
  \prpg{\psi_\alpha (y) \, \wbar{\psi}_{\dot{\beta}} (y')}(p)
     &=& p \!\cdot\! \sigma_{\alpha \dot{\beta}} \, G_{+-}(y,y',p;Z)  \,,\nn\\
  \prpg{\wbar{\psi}^{c\dot{\alpha}} (y) \, \psi^{c\beta} (y')}(p)
     &=& p \!\cdot\! \bar{\sigma}^{\dot{\alpha} \beta} \, G_{-+}(y,y',p;Z)  \,,
\eql{B-prop-no-flip}
\eeq
and
\beq
  \prpg{\psi_\alpha (y) \psi^{c\beta} (y')}(p)
     &=& \delta_\alpha^\beta \, \del_y G_{-+}(y,y',p;Z)  \,,\nn\\
  \prpg{\wbar{\psi}^{c\dot{\alpha}} (y) \, \wbar{\psi}_{\dot{\beta}} (y')}(p)
     &=& \!\!-\delta^{\dot{\alpha}}_{\dot{\beta}} \, \del_y G_{+-}(y,y',p;Z)
\eql{B-prop-chiral-flip}
\eeq
%

\section{Computation of the Soft Masses}

\subsection{Gauge Contributions}

We ignore the brane-localized kinetic terms for the gauge fields for
simplicity, as they have little relevance to the phenomenology we are
concerned with in this paper. The relevant bulk gauge interactions
involving the zero mode of the $B$ scalar are
\beq
  {\cal L}_{\text{bulk}} 
  &=& -\sqrt2 g_{\rm 5D}
       \bigl( \phi_B^\dag\lambda\psi_B - \phi_B \lambda^c\psi_B^c \bigr)
         \nn\\
  &\supset& 
      -\frac{\sqrt2 g}{\sqrt{1+Z}}
       \bigl( \phi_B^{(0)\dag}\lambda\psi_B 
             -\phi_B^{(0)} \lambda^c\psi_B^c \bigr)  \,,
\eeq
where the relation $g = g_{\rm 5D}/\sqrt{\pi R}$ was used in the second
line.

Then, the one-loop contribution to the squared soft mass of
$\phi_B^{(0)}$ from bulk fermion loops is then given by
\beq
 && -\frac{2ig^2 C_2}{1+Z} \int\!dy \int\!dy'\! \int\!\!\frac{d^4p}{(2\pi)^4}
        \nn\\
 &&  \hspace{1EM}
     \biggl(-{\rm tr}[p\!\cdot\!\sigma\, p\!\cdot\!\bar{\sigma}]
              \Bigl[ G_{+-}(y,y',p;Z) \, G_{+-}(y,y',p;0)
                    +G_{-+}(y,y',p;Z) \, G_{-+}(y,y',p;0)
              \Bigr]  \nn\\
 &&  \hspace{1.7EM}
            +{\rm tr}[{\mathbbm 1}] 
             \Bigl[ \del_y G_{+-}(y,y',p;Z) \, \del_y G_{+-}(y,y',p;0)   
                   +\del_y G_{-+}(y,y',p;Z) \, \del_y G_{-+}(y,y',p;0)
             \Bigr]            
     \biggr)  \nn\\
 &=& \frac{4g^2 C_2}{1+Z} \int\!dy \int\!dy'\! \int\!\frac{d^4p_E}{(2\pi)^4}
         \nn\\
 &&  \hspace{1EM}
     \biggl(p_E^2 \Bigl[ G_{+-}(y,y',p;Z) \, G_{+-}(y,y',p;0)
                        +G_{-+}(y,y',p;Z) \, G_{-+}(y,y',p;0)
                  \Bigr]  \nn\\
 &&  \hspace{2EM}
            +\del_y G_{+-}(y,y',p;Z) \, \del_y G_{+-}(y,y',p;0)   
            +\del_y G_{-+}(y,y',p;Z) \, \del_y G_{-+}(y,y',p;0)            
     \biggr) \,,
\eql{gauge-fermion-loop}
\eeq
where $p_E \equiv (-p^2-i\varep)^{1/2}$.  The first term is from the
loop of $\psi$ and $\lambda$, while the second from $\psi^c$ and
$\lambda^c$.  These do not contain chirality flips, and therefore the
propagators \eq{B-prop-no-flip} have been used.  The third and forth
terms come from the diagrams containing chirality flips, thus the
propagators \eq{B-prop-chiral-flip} have been used.

The bosonic contribution can be calculated by the following trick.
Imagine changing the boundary conditions for the fermions such that
supersymmety is preserved. In this situation, we know that the bosonic
and fermionic contributions cancel with each other.  Since we did not
change the boundary conditions for the bosons when switching from
non-supersymmetric to supersymmetric case, the bosonic contribution
stays the same. Thus, the bosonic contribution in the case of our
interest is just the negative of the fermionic contribution in the
supersymmetric case, i.e.,
\beq
  -\text{\eq{gauge-fermion-loop} with ``$+-$'' $\to$ ``$++$'' 
         and ``$-+$'' $\to$ ``$--$''.}
\eql{gauge-boson-loop}
\eeq
Adding \eq{gauge-fermion-loop} and \eq{gauge-boson-loop} gives the
squared soft mass from the bulk gauge interactions.

We also have contributions from the boundary-localized gauge interactions
\beq
  {\cal L}_{\text{boundary}} 
  &=& -Z\pi R\, \sqrt2 g_{\rm 5D} \,\phi_B^\dag\lambda\psi_B \bigr|_{y=0}  
          \nn\\
  &\supset& 
      -\frac{\sqrt2 g Z\pi R}{\sqrt{1+Z}} \phi_B^{(0)\dag}\lambda\psi_B
       \bigr|_{y=0}  \,.
\eeq
Using the above trick to obtain the bosonic part, this gives the
following contribution to the squared soft mass:
\beq
  && \frac{4g^2 (Z\pi R)^2 C_2}{1+Z} 
     \int\!\frac{d^4p_E}{(2\pi)^4}
      p_E^2 \Bigl[ G_{+-}(0,0,p;Z) \, G_{+-}(0,0,p;0)  \nn\\
  && \hspace{11.5EM} 
                 -G_{++}(0,0,p;Z) \, G_{++}(0,0,p;0)
           \Bigr]  \,,
\eql{gauge-brane-loop}
\eeq
Adding up \eq{gauge-fermion-loop}, \eq{gauge-boson-loop} and
\eq{gauge-brane-loop}, we obtain the formula \eq{softmass-gauge}.

\subsection{Yukawa Contributions}

The relevant Yukawa couplings are
\beq
  {\cal L}_{\text{yukawa}} 
  &=& -y_{\rm 5D} \, \tilde{h}_u 
       ( q_{3B} \tilde{u}_{3B} + \tilde{q}_{3B} u_{3B} )  \nn\\
  &\supset&
      -y_t \sqrt{\pi R(1+Z_{Q_3})} \, \tilde{h}_u q_{3B} \tilde{u}_{3B}^{(0)}
      -y_t \sqrt{\pi R(1+Z_{U_3})} \, \tilde{h}_u \tilde{q}_{3B}^{(0)} u_{3B}
\eeq
where
\beq
  y_{\rm 5D} = y_t \sqrt{\pi R(1+Z_{Q_3})} \sqrt{\pi R(1+Z_{U_3})}  \,.
\eeq
Then, again using the above trick to obtain the bosonic contribution, we
get
\beq
  \delta m_{\tilde{q}_3}^2
  &=& y_t^2 \pi R (1+Z_{U_3}) 
      \int\!\! \frac{d^4p}{(2\pi)^4}
        \biggl( -\frac{{\rm tr}[p\!\cdot\!\sigma\, p\!\cdot\!\bar{\sigma}]}
                      {p^2-\mu^2}
                 \Bigl[ G_{+-}(0,0,p;Z_{U_3}) - G_{++}(0,0,p;Z_{U_3}) \Bigr]
        \biggr)  \nn\\
  &\simeq&
     2y_t^2 \pi R (1+Z_{U_3}) 
     \int\!\! \frac{d^4p_E}{(2\pi)^4}
       \Bigl[-i G_{+-}(0,0,p;Z_{U_3}) +i G_{++}(0,0,p;Z_{U_3}) \Bigr]  \,,
\eeq
where $\mu$ was neglected in the second step.  This is the first
equation of \eq{softmass-Yukawa}.  Similarly, we have
\beq
  \delta m_{\tilde{u}_3}^2
  \simeq
     4y_t^2 \pi R (1+Z_{Q_3}) 
     \int\!\! \frac{d^4p_E}{(2\pi)^4}
       \Bigl[-i G_{+-}(0,0,p;Z_{Q_3}) +i G_{++}(0,0,p;Z_{Q_3}) \Bigr]  \,,
\eeq
which is the second equation of \eq{softmass-Yukawa}.



\begin{thebibliography}{99}

\bibitem{Little1}
 N.~Arkani-Hamed, A.~G.~Cohen and H.~Georgi,
  Phys.\ Lett.\  B {\bf 513}, 232 (2001)
  [arXiv:hep-ph/0105239].

\bibitem{Little2}
  N.~Arkani-Hamed, A.~G.~Cohen, E.~Katz, A.~E.~Nelson, T.~Gregoire and J.~G.~Wacker,
  JHEP {\bf 0208}, 021 (2002)
  [arXiv:hep-ph/0206020];
  N.~Arkani-Hamed, A.~G.~Cohen, E.~Katz and A.~E.~Nelson,
  JHEP {\bf 0207}, 034 (2002)
  [arXiv:hep-ph/0206021];
  T.~Gregoire and J.~G.~Wacker,
  JHEP {\bf 0208}, 019 (2002)
  [arXiv:hep-ph/0206023];
  I.~Low, W.~Skiba and D.~Tucker-Smith,
  Phys.\ Rev.\  D {\bf 66}, 072001 (2002)
  [arXiv:hep-ph/0207243];
  D.~E.~Kaplan and M.~Schmaltz,
  JHEP {\bf 0310}, 039 (2003)
  [arXiv:hep-ph/0302049].

\bibitem{twin}
  Z.~Chacko, H.~S.~Goh and R.~Harnik,
  Phys.\ Rev.\ Lett.\  {\bf 96}, 231802 (2006)
  [arXiv:hep-ph/0506256];
  R.~Barbieri, T.~Gregoire and L.~J.~Hall,
  arXiv:hep-ph/0509242;
  Z.~Chacko, Y.~Nomura, M.~Papucci and G.~Perez,
  JHEP {\bf 0601}, 126 (2006)
  [arXiv:hep-ph/0510273].

\bibitem{twinLR}
  Z.~Chacko, H.~S.~Goh and R.~Harnik,
  JHEP {\bf 0601}, 108 (2006)   
  [arXiv:hep-ph/0512088];
 H.~S.~Goh and C.~A.~Krenke,
  Phys.\ Rev.\  D {\bf 76}, 115018 (2007)
  [arXiv:0707.3650 [hep-ph]].
  
\bibitem{folded}
  G.~Burdman, Z.~Chacko, H.~S.~Goh and R.~Harnik,
  JHEP {\bf 0702}, 009 (2007)
  [arXiv:hep-ph/0609152].

\bibitem{BCGHK}
  G.~Burdman, Z.~Chacko, H.~S.~Goh, R.~Harnik and C.~A.~Krenke,
  arXiv:0805.4667 [hep-ph].

\bibitem{markus}
  J.~Kang and M.~A.~Luty,
  arXiv:0805.4642 [hep-ph].

\bibitem{bounds1}
  P.~F.~Smith and J.~R.~J.~Bennett,
  Nucl.\ Phys.\  B {\bf 149}, 525 (1979);
  C.~B.~Dover, T.~K.~Gaisser and G.~Steigman,
  Phys.\ Rev.\ Lett.\  {\bf 42}, 1117 (1979);
  P.~F.~Smith, J.~R.~J.~Bennett, G.~J.~Homer, J.~D.~Lewin, H.~E.~Walford and W.~A.~Smith,
  Nucl.\ Phys.\  B {\bf 206}, 333 (1982).


\bibitem{bounds2}
  T.~K.~Hemmick {\it et al.},
  Phys.\ Rev.\  D {\bf 41}, 2074 (1990);
  P.~Verkerk, G.~Grynberg, B.~Pichard, M.~Spiro, S.~Zylberajch, M.~E.~Goldberg and P.~Fayet,
  Phys.\ Rev.\ Lett.\  {\bf 68}, 1116 (1992);
  T.~Yamagata, Y.~Takamori and H.~Utsunomiya,
  Phys.\ Rev.\  D {\bf 47}, 1231 (1993);
  R.~N.~Mohapatra and S.~Nussinov,
  Phys.\ Rev.\  D {\bf 57}, 1940 (1998)
  [arXiv:hep-ph/9708497].

\bibitem{bounds3}
  M.~Byrne, C.~F.~Kolda and P.~Regan,
  Phys.\ Rev.\  D {\bf 66}, 075007 (2002)
  [arXiv:hep-ph/0202252].


\bibitem{SS}
  J.~Scherk and J.~H.~Schwarz,
  Phys.\ Lett.\  B {\bf 82}, 60 (1979);
  J.~Scherk and J.~H.~Schwarz,
  Nucl.\ Phys.\  B {\bf 153}, 61 (1979).


\bibitem{PQ}
  A.~Pomarol and M.~Quiros,
  Phys.\ Lett.\  B {\bf 438}, 255 (1998)
  [arXiv:hep-ph/9806263];
  I.~Antoniadis, S.~Dimopoulos, A.~Pomarol and M.~Quiros,
  Nucl.\ Phys.\  B {\bf 544}, 503 (1999)
  [arXiv:hep-ph/9810410];
  A.~Delgado, A.~Pomarol and M.~Quiros,
  Phys.\ Rev.\  D {\bf 60}, 095008 (1999)
  [arXiv:hep-ph/9812489].

\bibitem{BHN}
  R.~Barbieri, L.~J.~Hall and Y.~Nomura,
  Phys.\ Rev.\  D {\bf 63}, 105007 (2001)
  [arXiv:hep-ph/0011311].

\bibitem{Take}
  L.~J.~Hall, Y.~Nomura, T.~Okui and S.~J.~Oliver,
  Nucl.\ Phys.\  B {\bf 677}, 87 (2004)
  [arXiv:hep-th/0302192];
  J.~Bagger and D.~V.~Belyaev,
  JHEP {\bf 0306}, 013 (2003)
  [arXiv:hep-th/0306063].

\bibitem{pedestrian}
  Z.~Chacko, Y.~Nomura and D.~Tucker-Smith,
  Nucl.\ Phys.\  B {\bf 725}, 207 (2005)
  [arXiv:hep-ph/0504095].
  
\bibitem{DX}
  M.~Drees and X.~Tata,
  Phys.\ Lett.\  B {\bf 252}, 695 (1990).

\bibitem{DDRT}
  S.~Dimopoulos, M.~Dine, S.~Raby and S.~D.~Thomas,
  Phys.\ Rev.\ Lett.\  {\bf 76}, 3494 (1996)
  [arXiv:hep-ph/9601367];
  S.~Ambrosanio, G.~D.~Kribs and S.~P.~Martin,
  Phys.\ Rev.\  D {\bf 56}, 1761 (1997)
  [arXiv:hep-ph/9703211].

\bibitem{ADPRW}
  A.~Arvanitaki, S.~Dimopoulos, A.~Pierce, S.~Rajendran and J.~G.~Wacker,
  Phys.\ Rev.\  D {\bf 76}, 055007 (2007)
  [arXiv:hep-ph/0506242].

\bibitem{splitSUSY}
  N.~Arkani-Hamed and S.~Dimopoulos,
  JHEP {\bf 0506}, 073 (2005)
  [arXiv:hep-th/0405159];
  G.~F.~Giudice and A.~Romanino,
  Nucl.\ Phys.\  B {\bf 699}, 65 (2004)
  [Erratum-ibid.\  B {\bf 706}, 65 (2005)]
  [arXiv:hep-ph/0406088].

\end{thebibliography}
\end{document}